\documentclass[prl,twocolumn,amsmath,amssymb]{revtex4}

\usepackage{amsmath}
\usepackage{amssymb}
\usepackage{graphicx}
\usepackage{upgreek}

\makeatletter

\newcommand{\Rmnum}[1]{\expandafter\@slowromancap\romannumeral
#1@}
\makeatother

\usepackage{float}

\usepackage[T1]{fontenc}
\usepackage{indentfirst}
\begin{document}

\title{Anomalous Antiferromagnetism in Metallic RuO$_2$ Determined by Resonant X-ray Scattering}
\author{Z. H. Zhu$^{1*}$, J. Strempfer$^2$, R. R. Rao$^3$, C. A. Occhialini$^1$, J. Pelliciari$^1$, Y. Choi$^2$, T. Kawaguchi$^4$, H. You$^4$, J. F. Mitchell$^4$, Y. Shao-Horn$^{3, 5}$, and R. Comin$^1$ }
\email[Correspondence should be addressed to ]{zzh@mit.edu;rcomin@mit.edu (R. Comin)}
\affiliation{$^1$Department of Physics, Massachusetts Institute of Technology, Cambridge, MA 02139, USA\\$^2$Advanced Photon Source, Argonne National Laboratory, Argonne, Illinois 60439, USA\\ $^3$Department of Mechanical Engineering, Massachusetts Institute of Technology, Cambridge, MA 02139, USA \\$^4$Materials Science Division, Argonne National Laboratory, Argonne, Illinois 60439, USA\\$^5$Department of Material Science and Engineering, Massachusetts Institute of Technology, Cambridge, MA 02139, USA}

\date{\today}

\begin{abstract}We studied the magnetic ordering of thin films and bulk crystals of rutile RuO$_2$ using resonant X-ray scattering across the Ru L$_2$ absorption edge. Combining polarization analysis and azimuthal-angle dependence of the magnetic Bragg signal, we have established the presence and characteristic of collinear antiferromagnetism in RuO$_2$ with T$_N$ $>$ 300 K. In addition to revealing a spin-ordered ground state in the simplest ruthenium oxide compound, the persistence of magnetic order even in nanometer-thick films lays the ground for potential applications of RuO$_2$ in antiferromagnetic spintronics.
\end{abstract}
\maketitle
In electronic systems with localized $d$ electrons and an insulating ground state, ordered magnetism arises from strong exchange interactions that are often described within the framework of the Heisenberg model. However, in metals with partly itinerant $d$ electrons, it is often more appropriate to interpret magnetic phenomena  on the basis of correlation effects between band-like states. A basic understanding of magnetism in ferromagnetic metals has been obtained at a level of mean field approximation and beyond, in the framework of the Hubbard model~\cite{moriya_recent_1979}. However, a general description of spin order in antiferromagnetic metals remains challenging. The best-known example is probably that of Cr metal, whose incommensurate spin density wave (SDW) is characterized by a wave vector determined by the nesting properties of its Fermi surface ~\cite{fawcett_spin_1988}. Some perovskite chromates, such as CaCrO$_3$ and SrCrO$_3$, have been recently established as antiferromagnetic metals (AFMs) as well, but the roots of the AFM order has remained elusive ~\cite{komarek_magnetic_2011,zhou_anomalous_2006,williams_charge_2006,ortega-san-martin_microstrain_2007,komarek_$mathrmcacro_3$:_2008,streltsov_band_2008,bhobe_electronic_2011}.

The family of ruthenium based perovskite oxides encircles several compounds with a rich phenomenology and distinct electronic ground states. For example, Sr$_2$RuO$_4$ exhibits unconventional superconductivity with triplet pairing below 2 K ~\cite{maeno_superconductivity_1994} while its close structural relative Sr$_3$Ru$_2$O$_7$ has a metamagnetic ground state ~\cite{perry_metamagnetism_2001}. Ca$_2$RuO$_4$ ~\cite{nakatsuji_ca_1997} and Ca$_3$Ru$_2$O$_7$ ~\cite{yoshida_crystal_2005} are antiferromagnetic insulators in their ground sates, whereas CaRuO$_3$ and SrRuO$_3$ are a paramagnetic and ferromagnetic metal, respectively ~\cite{longo_magnetic_1968}. The parent compound RuO$_2$ was long assumed to be paramagnetic and metallic. This characterization was primarily based on measurements of bulk magnetization ~\cite{ryden_magnetic_1970,guthrie_magnetic_1931}. However, both a quadratic and linear temperature dependence of the magnetic susceptibility were observed in these studies, which have motivated further diffraction studies to resolve the microscopic spin structure of RuO$_2$. A recent neutron diffaction study on single crystal RuO$_2$ reported antiferromagnetic order up to at least 300 K with a small room temperature magnetic moment of approximately 0.05 $\mu_B$ ~\cite{berlijn_itinerant_2017-1}. This discovery does not only raise new inquiries into the nature of itinerant antiferromagnetism in RuO$_2$, but also underscores its potential use in antiferromagnetic spintronic devices, which are drawing considerable attention recently~\cite{zelezny_spin_2018,baltz_antiferromagnetic_2018,fukami_magnetization_2016,nunez_theory_2006,park_spin-valve-like_2011,tshitoyan_electrical_2015}. Ruthenium oxide possesses the key traits of a spintronic material: a metallic ground state; room temperature antiferromagnetism, with high N\'{e}el temperature $T_N$($>$ 300 K); and a theoretically proposed collinear AFM structure, which has not been fully resolved, to date. To assess the potential of RuO$_2$ for spintronic applications, it is also essential to establish the presence of magnetic order in thin film materials that can serve as a basis for the fabrication of electronic devices.

Here, we used resonant X-ray Scattering (RXS) measurements at the Ru L$_2$ resonance ($\approx$ 2.968 keV) to investigate antiferromagentic order in RuO$_2$ thin films. RXS is a photon-in/photon-out and element-specific probe of electronic orders in the Fourier domain, and has been previously used to detect spin ordering in other ruthenate compounds ~\cite{bohnenbuck_magnetic_2009,Bohnenbuck_magnetic_2008, zegkinoglou_orbital_2005,nelson_spin-charge-lattice_2007}. In particular, resonant diffraction is a well-suited probe of magnetism in thin film materials, whose thickness (1-100 nm) is ideally matched with the probing depth of tender X-ray photons (200-500 nm). In this study, we establish the presence of antiferromagnetism and resolve the underlying spin texture in thin films and bulk crystals of RuO$_2$.          

Thin film samples of (1 0 0) oriented RuO$_2$ have been synthesized on (0 0 1) SrTiO$_3$ substrates via pulsed laser deposition (PLD) using a KrF excimer laser ($\lambda$ = 248 nm) ~\cite{stoerzinger_orientation-dependent_2014}. Single crystals are grown by the chemical vapor transport (CVT) method. RuO$_2$ powder was heated to $\approx$ 1300 $^\circ$C at one end of the tube, while maintaining a linear flow of oxygen gas to transport the vaporized material. RuO$_2$ single crystals slowly form in the cooler zone at the opposite end of the tube ~\cite{lister_cathodic_2003}. The RXS experiments were performed at beam line 4-ID-D of the Advanced Photon Source at Argonne National Laboratory. The scattering measurements were conducted in vacuum to minimize beam attenuation using a windowless vacuum shroud on loan from beamline P09/PETRA \Rmnum{3} ~\cite{strempfer_resonant_2013}. The polarization analysis of scattered photons was carried out using a Si(111) analyzer crystal. 
\begin{figure}[htb]
\centering
\includegraphics[scale=1.0]{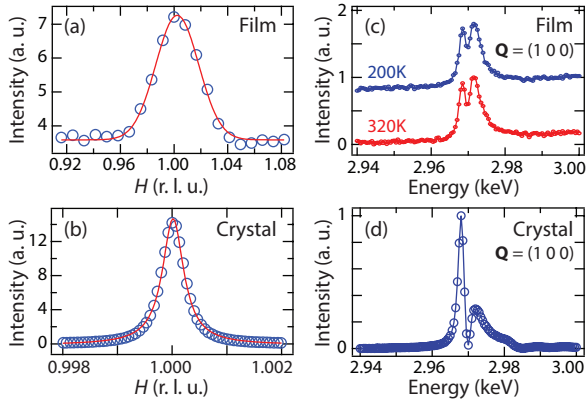}
\caption{\label{fig:epsart} Resonant enhancement of the (1 0 0) reflection at the Ru-$L_2$ edge. (A) and (B) show longitudinal reciprocal-space scans across wavevector Q$_{AFM}$ = (1 0 0) using a photon energy of 2.9685 keV for a typical thin film and bulk crystal, respectively. The solid lines in both plots represent Gaussian fits to data. (C) represents the energy dependence of the scattered intensity taken at the (1 0 0) reflection of the film for different temperatures; the measurements are offset for clarity. (D) shows the energy dependence of the scattered  intensity from the bulk crystal at Q$_{AFM}$ and 300 K. The solid lines in (C) and (D) are to guide the eye.}
\vspace{-0.0cm}
\end{figure}

In Figure 1(A) and (B) we show the longitudinal momentum scans across wavevector Q$_{AFM}$ = (1 0 0), which is a structurally forbidden reflection, at the Ru $L_2$ resonance (2.9685 keV) for a typical thin film and bulk crystal, respectively. To further demonstrate the electronic nature of the (1 0 0) reflection from the resonant enhancement of the scattering cross section, Figure 1(C) and (D) show the photon energy dependence of the scattered intensity for the film and bulk crystal, respectively. As shown in Figure 1(C), a double-peak structure is observed for the RXS intensity at fixed wavevector (1 0 0). The first peak at 2.9685 keV reflects the resonant enhancement of the scattering cross section, which arises from electric dipole transitions from 2p$_{1/2}$ core levels directly into the partially occupied 4d $t_{2g}$ orbitals. The second peak at 2.9716 keV is likely due to transitions into the unoccupied 4d $e_g$ orbitals, similar to those previously observed in Ca$_2$RuO$_4$ and RuSr$_2$GdCu$_2$O$_8$~\cite{bohnenbuck_magnetic_2009,zegkinoglou_orbital_2005,fang_orbital-dependent_2004}. As shown in Figure 1(D), the double-peak structure is similarly found in the bulk crystal, however the scattering resonance profile is also more asymmetric. This difference between the film and bulk crystal is due to the self-absorption effect in the bulk case. The film thickness (25 nm) is much shorter than the absorption length (600 nm at 2.9685 keV; see Supplemental Material for details ~\cite{supplemental, chantler_detailed_2000, chantler_theoretical_1995, haverkort_symmetry_2010}), and the entire film is probed within the energy range used. However, in the case of bulk crystal, the thickness is much larger than the absorption length, and the probing volume changes significantly with the incident energy. Especially, the higher energy peak in Figure 1(D) is located near the whiteline of the x-ray absorption, and this reduces the probing depth significantly, suppressing the peak intensity. Nevertheless, the equivalence in the ordering  wavevector and the similarity of the profiles of diffracted intensity vs. photon energy suggest that the observed resonant reflections for both samples originate from the same phenomenon, namely the magnetic order proposed in the previous neutron scattering study ~\cite{berlijn_itinerant_2017-1}.   

\begin{figure}[htb]
\begin{center}
\includegraphics [scale =1.0]{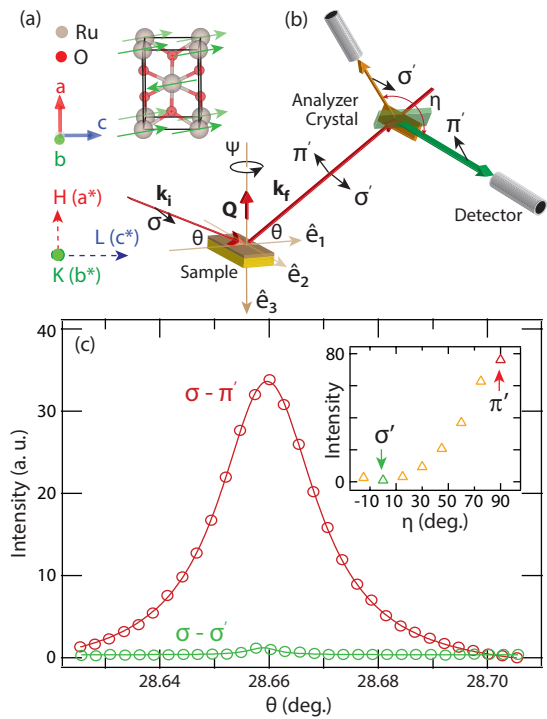}
\end {center}
\caption {\label{fig:epsart} Magnetic ordering from polarization dependence of the scattered intensity at the (1 0 0) reflection. (A) illustrates the proposed AFM structure for rutile RuO$_2$, constructed on a tetragonal unit cell (space group P42/mnm) with lattice parameters a $\approx$ 4.49 \AA \ and c $\approx$ 3.11 \AA. The light gray and red circles represent Ru and O atoms, respectively. (B) is a schematic plot of the scattering geometry. (C) shows the polarization dependence of the scattered intensity at Q = (1 0 0). The solid lines are fits to Gaussian profiles. The inset is the scattered intensity as a function of angle $\eta$; the $\sigma^{\prime}$ and $\pi^{\prime}$ beams are detected at $\eta$ = 0$^{\circ}$ and $\eta$ = 90$^{\circ}$, respectively.}
\vspace{0.0cm}
\end{figure}
 
In order to distinguish between magnetic and charge channels for the observed resonant reflection, we carried out photon polarization analysis using a Si(111) analyzer crystal. The polarization of the incoming photon is fixed to $\sigma$ and we measured the out-going photon in both  $\sigma^{\prime}$ and $\pi^{\prime}$ polarization projections of the scattered photons, where $\sigma$ and $\pi$ represent, respectively, the polarization component perpendicular and parallel to the scattering plane. Figure 2(A) shows the unit cell of rutile RuO$_2$ with the collinear AFM structure. The fundamental magnetic wavevector is (1 0 0) as we have observed at Ru-L$_2$ edge. Figure 2(B) graphically systematizes the experimental configuration. The selection between $\sigma$-$\pi^{\prime}$  and $\sigma$-$\sigma^{\prime}$ is controlled by rotating the analyzer around the scattered wavevector $K_f$ by $\eta$ = 90$^\circ$. As shown in Figure 2(C), the intensity of the rocking curve at the (1 0 0) position is dominant in the $\sigma$-$\pi^{\prime}$ channel. The inset in Figure 2(C) shows the integrated intensity of the analyzer scans as a function of $\eta$, which reaches its maximum in the $\sigma$-$\pi^{\prime}$ channel and near zero in the $\sigma$-$\sigma^{\prime}$ channel. 
\begin{figure}[htp]\includegraphics[scale = 1]{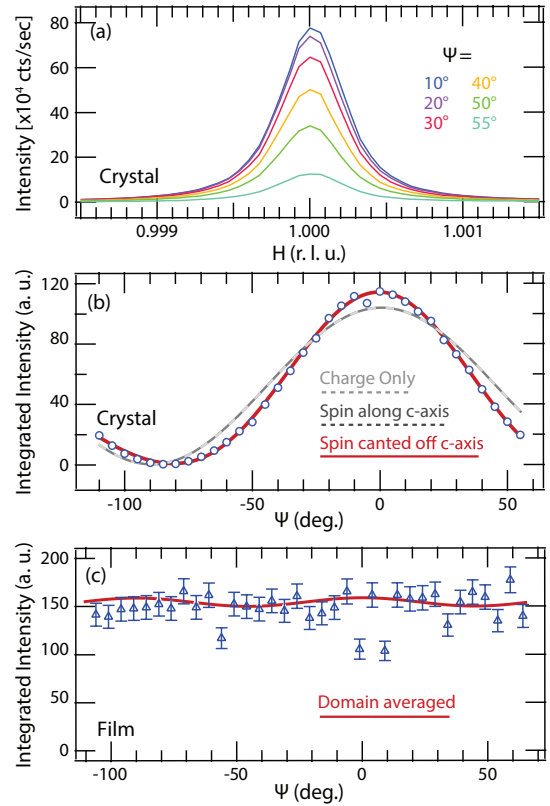}
\centering
\caption{\label{fig:epsart} (A) The scattered intensity of the magnetic reflection (1 0 0) for various azimuthal angles $\Psi$ at 300 K. (B) Azimuthal dependence of the integrated intensity from fits to the data set in (A) using a Gaussian profile at the reflection (1 0 0) at 300 K. (where not shown, the error bars are within the symbol size.) The azimuthal angle $\Psi$ = 0 corresponds to a sample orientation where the c-axis lies nearly within the scattering plane. The black and gray lines represent the best fit for a model containing symmetry allowed charge-anisotropy or first-order magnetic scattering with moments oriented strictly along the c-axis, while the solid red line corresponds to a best fit including terms up to second order in the magnetic scattering process, and canting of the spin moment away from the c-axis as explained in the text. (C) Azimuthal dependence of the integrated intensity of the magnetic reflection (1 0 0) for a typical film sample and the solid lines represent fits, as explained in the text. The discrepancy between the bulk and film samples arise from the presence of multiple domains in the latter. } 
\end{figure}

To decode the origin of the resonant reflection at the wave vector (1 0 0), we carried out a detailed symmetry-restricted tensorial analysis of the azimuthal angle dependence of the scatterred intensity (see Supplemental Material for details~\cite{supplemental}). In Figure 3(A), we show a series of representative momentum scans across the (1 0 0) reflection for different azimuthal angles $\Psi$. In Figure 3(B), we plot the integrated peak area (extracted from Gaussian fits to the momentum scans) as a function of $\Psi$. A clear modulation with a period of $\pi$ can be visually inferred, with the magnetic scattering intensity being maximized at $\Psi$ = 0$^\circ$ and minimized at $\Psi$ = -90$^\circ$, where the azimuthal angle  $\Psi$ = 0$^\circ$ corresponds to the (0 0 1) direction lying in the diffraction plane. A pure two-fold modulation at the (1 0 0) reflection can be equivalently described by a model derived by scattering from quadrupolar-type charge anisotropy or magnetic scattering with moments oriented strictly along the c-axis (see Supplementary Material for further details and a more extended description of the model~\cite{supplemental}).  The best-fit result for a pure scattering of the type $I_{\sigma\pi} \propto |\cos(\Psi)|^2$ corresponding to these two distinct mechanisms is given in Figure 3(B) as the gray and black dotted lines.  In the charge anisotropy picture, this is the only component that is allowed by symmetry of the Ru atoms in the $P_4/mnm$ space group. However, in the case that the scattering is of magnetic origin, one can generalize the model to include a slight canting of the moment off of the c-axis. The resulting higher harmonic content in the azimuthal dependence of the scattering intensity arises when more than a single component of the magnetic moment is nonzero. A fit to this generalized model is given by the solid red curve in Figure 3(B), showing a significantly improved agreement to the data, further supporting the magnetic origin of this peak. We caution, however, that with only a single observable reflection at Ru-L$_2$ edge, it is not possible to completely rule out a partial contribution from charge anisotropy. Figure 3(C) reports the azimuthal dependence of the magnetic scattering intensity from the thin film sample. Unlike the case of the bulk crystal, it does not exhibit any significant modulation with $\Psi$. This seeming discrepancy between the thin film and bulk crystal is explained by the twinned nature of the thin films, reflecting the existence of multiple domains. The introduction of an arbitrarily canted moment off of the high-symmetry axis necessitates the consideration of eight distinct species of orthogonal domains sharing a common (1 0 0) epitaxial axis either parallel or antiparallel to the scattering vector. The local scattering intensity from simultaneously probed domains can be related to a global azimuthal angle $\Psi$ by a domain-dependent phase shift and rotations by $\pm$ $\Psi$.  Taking the best-fit parameters from the bulk case in Figure 3(B), incoherently averaging the contributions and using the known orientation of the film at zero azimuth yields the predicted dependence for the magnetic peak in the thin film as shown in  Figure 3(C). The domain-averaging has suppressed lower-frequency components of the azimuthal dependence, leaving a nearly constant intensity with a small residual higher-frequency modulation.  The amplitude of the residual modulation lies well below the noise and the prediction is consistent with the observations within experimental uncertainty.

\begin{figure}[htp]\includegraphics[scale=1.00]{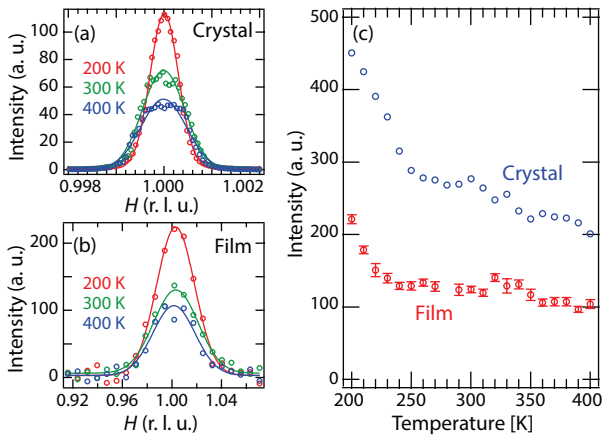}
\centering
\caption{\label{fig:epsart}The scattered intensity of the resonant reflection (1 0 0) at various temperatures for a typical film (A) and bulk crystal (B), respectively. The solid lines are fits to Gaussian profiles. (C) The temperature dependence of the scattered intensity of the resonant reflection (1 0 0) for a  typical film and bulk crystal, respectively. } 
\vspace{0.0cm}
\end{figure}

A major aspect and motivation of our investigation is to assess whether magnetic order persists up to room temperature in RuO$_2$ films. Figure 4(A) and (B) show representative longitudinal scans across the magnetic ordering vector at various temperatures, for the bulk crystal and thin film, respectively. For both samples, the magnetic scattering yield decreases when increasing temperature but spin order persists up to at least 400 K. The temperature dependence of the scattered intensity is shown in Figure 4(C) (data are rescaled and offset for clarity). Both samples exhibit a remarkably similar temperature evolution of the magnetic order parameter. The intensity diminishes rapidly, a factor of 2 from 200 K to 250 K, then smoothly tails off above 250 K and up to 400 K. The increase in the line-width of the (1 0 0) reflection for increasing temperatures suggests a progressive reduction in the spin-spin correlation lengths, which however remain finite across the whole temperature range surveyed in this study. In any case, the energy dependence of the magnetic scattering in the thin film sample is almost identical at 200 K and 320 K (see again Figure 1(C)), indicating that the emergence of a second phase below 250 K is unlikely. The observed resonant reflection at Q = (1 0 0) retains its magnetic character both below and above 250 K within the range of measured temperatures.

In summary, we have established the existence of collinear antiferromagnetism in both thin film and bulk rutile RuO$_2$. At odds with previous reports that AFM order is short-ranged [15], our experiments reveal sharp antiferromagnetic diffraction signatures in bulk crystals with a correlation length in excess of 4000 \AA\ in the $a$ direction. In the thin film, the correlation length is reduced to about 50 \AA \ in the $a$ direction. This far smaller correlation length compared to the bulk crystal can be accounted for by the dimensional confinement along the $a$ direction in the thin film. The azimuthal angle dependence analysis suggests a collinear AFM magnetic structure with spin moments having dominant projection along (0 0 1), which agrees well with theoretical predictions from density functional theory ~\cite{berlijn_itinerant_2017-1}. The emergence of antiferromagnetism in a highly conducting oxide is rare and unusual, often implying some exotic physics at play. The AFM instability manifested by RuO$_2$ may evoke some analogies with the paradigmatic case of Cr metal, a spin-density-wave antiferromagnet. However, and at variance with the incommensurate AFM ordering of Cr metal, here the magnetic diffraction data rule out any significant incommensurablity of the magnetic wave vector in RuO$_2$. This fact might suggest a possibly different origin of the observed AFM spin textures. The itinerant AFM state of RuO$_2$ is also reminiscent of the anomalous magnetism found in some perovskite chromates (CaCrO$_3$ and SrCrO$_3$), whose origin has long remained unclear. Our experiments in highly conducting RuO$_2$ thus raise fundamental inquiry into the nature of the itinerant antiferromagnetism in this 4$d$ transition metal oxide. From an applied perspective, the presence of room temperature antiferromagnetism in 25-nm films of a metallic oxide underscore RuO$_2$ a potential candiate for spintronic devices. In addition, the evidence of the magnetic moments in RuO$_2$ may prove to be important in catalysis of oxygen evolution reaction ~\cite{r.rao_towards_2017} where the spin conservation rule plays an important role in producing oxygen molecules with spin~\cite{torun_role_2013}.

This research was supported by NSF through the Massachusetts Institute of Technology Materials Research Science and Engineering Center DMR - 1419807. R. C. acknowledges support from the Alfred P. Sloan Foundation. J. P. acknowledges financial support by the Swiss National Science Foundation Early Postdoc.Mobility fellowship project number P2FRP2$\_$171824. The work in the Materials Science Division of Argonne National Laboratory (bulk crystal synthesis) was supported by the U.S. Department of Energy, Office of Science, Basic Energy Sciences, Materials Science and Engineering Division. The work performed at the Advanced Photon Source was supported by the U.S. Department of Energy, Office of Science, and Office of Basic Energy Sciences under Contract No. DE-AC02-06CH11357 

\scriptsize

\bibliography{RuO2_mainTEXT}

\end{document}